\documentclass[a4paper,fleqn,usenatbib]{mnras}

\usepackage{newtxtext,newtxmath}

\usepackage[T1]{fontenc}
\usepackage{ae,aecompl}

\usepackage{graphicx}	
\usepackage{amsmath}	
\usepackage{amssymb}	
\usepackage{natbib}

\title{Kinematics and Structure of Star-forming Regions: Insights from Cold Collapse Models}

\author[Kuznetsova et al.]{
Aleksandra Kuznetsova,$^{1}$\thanks{E-mail: kuza@umich.edu}
Lee Hartmann,$^{1}$
Javier Ballesteros-Paredes$^{2,3}$
\\
$^{1}$Department of Astronomy, University of Michigan, 1085 S. University Ave., Ann Arbor, MI 48109\\
$^{2}$Centro de Radioastronom\'ia y Astrof\'isica, Universidad Nacional Aut\'onoma de M\'exico,
            Apdo. Postal 72-3 (Xangari), \\ Morelia, Michoc\'an 58089, M\'exico\\
$^{3}$ Zentrum fur Astronomie der Universitat Heidelberg, Institut fur Theoretische Astrophysik, Albert-Ueberle-Str. 2, \\ 69120 Heidelberg, Germany
}

\date{Accepted XXX. Received YYY; in original form ZZZ}

\pubyear{2017}

\begin{document}
\label{firstpage}
\pagerange{\pageref{firstpage}--\pageref{lastpage}}
\maketitle

\begin{abstract}
The origin of the observed morphological and kinematic substructure of young star forming regions is a matter of debate.  We offer a new analysis of data from simulations of globally gravitationally collapsing clouds of progenitor gas to answer questions about sub-structured star formation in the context of cold collapse. As a specific example, we compare our models to recent radial velocity survey data from the IN-SYNC survey of Orion
and new observations of dense gas kinematics, and offer possible interpretations of kinematic and morphological signatures in the region. In the context of our model, we find the frequently-observed hub-filament morphology of the gas naturally arises during gravitational evolution, as well as the dynamically-distinct kinematic substructure of stars.  We emphasize that the global and not just the local gravitational potential plays an important role in determining the dynamics of both clusters and filaments. 
\end{abstract}


\section{Introduction}
\par Star forming regions generally show significant substructure on a variety of scales. Observations of the dust continuum have found that many star forming regions are threaded by networks of filaments, often connected to ``hubs" of dense gas and clusters of stars \citep{Myers_2009, Andre_2010}. Studies of the dense gas in these filaments have identified fibrous substructure by correlating related structures in position-position-velocity space \citep{Hacar_2013}. 
It has become clear that models of star formation must be able to explain the origin and role of filaments, clumps, cores, and related substructures. 
\par Previous work has looked at filament formation as the result of the collisions of turbulent planar shocks \citep[e.g.][]{MacLow_2004, Federrath_2016} and even cloud-cloud collisions
 \citep{Balfour_2015}.
In this paradigm, the formation of star clusters occurs at the intersections of filaments, where there is enough dense material collected. However, globally collapsing non-spherical mass distributions can also form filaments  because of collapse along the shortest axis first. 
\citet{Hartmann_2007} applied this idea to a collapsing, rotating elliptical sheet of gas and were able to qualitatively reproduce the structure of the Orion A cloud, in particular the structure of the integral-shaped filament of dense gas running through the Orion Nebula Cluster (ONC).

Multi-object high resolution spectrometers have begun to add to the understanding of star-forming regions with large surveys investigating the kinematic as well as the spatial relationships of the young stars in clusters to the dense gas \citep{Tobin_2009,Kounkel_2016}.  The most recent of these, from the IN-SYNC near-infrared spectroscopic survey, has
provided the clearest analysis yet of kinematic and spatial properties in Orion A  \citep{DaRio_2017}.  This investigation showed that (a) the stellar radial velocities mostly follow the nearby molecular gas motions; (b) there is a small blueshift of the stars relative to the gas, as originally found in the optical survey of \citet{Tobin_2009}; (c) several groups exhibit
kinematic subclustering, with the Orion ONC showing a smoother velocity structure; and (d) there is evidence
that some of the groups may be gravitationally unbound, with the ONC itself possibly expanding.  On the other hand, \citet{Hacar_2017} infer that the dense gas in the OMC-1 cloud associated with the ONC is
gravitationally collapsing, based on observations of N$_2$H$^+$.

These new results on gas and stellar kinematics motivated us to perform a new analysis of simulations from \citet{Kuznetsova_2015} to characterize the substructure expected from a young star forming region formed by global gravitational collapse.  Our simulations assume ``fast" star formation, where initial turbulence seeds the cloud with density perturbations; the subsequent evolution is driven by gravity without continued turbulent
forcing.  While stellar feedback (jets, winds, photoionzation/dissociation) can be important and even dominant once massive stars form, this simplification allows us to identify features 
which can be explained by gravitational collapse
alone. \citet{Kuznetsova_2015} found that cold collapse models can naturally produce a cluster over short timescales, but that the rapid dynamical evolution inherent in global gravitational collapse tends to erase obvious kinematic signatures of infall within cluster bounds. In this new analysis, we find that several of the features
identified by \citet{DaRio_2017} in Orion A
can be explained qualitatively and semi-quantitatively by gravitational driving.
Our results are also in good agreement with the inference of gravitational collapse by \citet{Hacar_2017}.

\section{Method}

\par We analyse data from a simulation of the sub-virial collapse of a 2320 $M_{\odot}$ triaxial ellipsoidal shell of molecular gas (run LR from  \citet{Kuznetsova_2015}) using the smoothed-particle-hydrodynamics (SPH) code Gadget2 \citep{Springel_2005} with a sink implementation from \citet{Jappsen_2005}. As in \citet{Hartmann_2007}, the cloud is given some element of rotation along the long axis. Supersonic turbulence was seeded initially with the same prescription as in \citet{Ballesteros-Paredes_2015}.  Because no
continued forcing is applied,
turbulent velocities quickly damp, leaving behind small-scale density fluctuations which collapse to form sinks while the overall cloud evolves purely under gravity.
The run was evolved for $1.1$ free-fall times, where $t\mathrm{_{ff}} = 32\pi (G\rho)^{-1/2} = 0.85$ Myr. 

\par To make a comparison between our simulations and the kinematic substructure observed by \citet{DaRio_2017}, we use an algorithm based on friends-of-friends in order to correlate substructure spatially and kinematically. Friends-of-friends (FOF) is a clustering algorithm, commonly used for identifying halos in cosmological simulations \citep{Huchra_1982}, that finds groups of members that are denser than a threshold value. For our dataset, we use the 3-D spatial data to locate groups of sinks in which all members are at a linking length, $b$, away from one another. We use a linking length that corresponds to an overdensity of $n = 125 n_0$, where $n_0$ is the number density of stars at $t = 1.1 t\mathrm{_{ff}}$ if they were homogeneously distributed across the initial volume.  To ensure that the groups found are a single moving group, i.e. a group with a characteristic mean velocity and a given velocity dispersion, we further refine the spatially identified clusters by first making sure that the distribution in velocity space peaks around a single value. However, we merge groups that both a) could be grouped as one structure with a linking length on the order of the structures' sizes and b) have velocity dispersions larger than their difference in mean velocities. Then, we iterate over all groups with a relaxed linking length criterion and accept new members if the velocities of potential members are within 1 standard deviation of the mean velocity of the group. 

\par  Because proper motions are generally not available, similar analyses on observational data must either locate groups by identifying overdensities in position-position-velocity (PPV) space \citep{DaRio_2017} or by employing an FOF-like scheme in PPV space \citep{Hacar_2013}. \citet{DaRio_2017} create a stellar density field by sampling the local density at regular points in PPV space, using the closest $n$ stars and a conversion
metric such that $1^{\circ}$ corresponds to
$4 \, \mathrm{km s^{-1}}$. Structures are then identified as local maxima in the field. 

While PPV methods identify structures in our simulation that are qualitatively similar to those identified with our 3-D FOF scheme, the group properties are much less robust in the former than the latter, mainly because we have about a factor of ten fewer sink particles than stars with radial velocities in the IN-SYNC survey. Therefore we opt to characterize the substructures from the 3-D kinematically refined FOF scheme.

\begin{figure}
    \centering
    \includegraphics[width = \columnwidth]{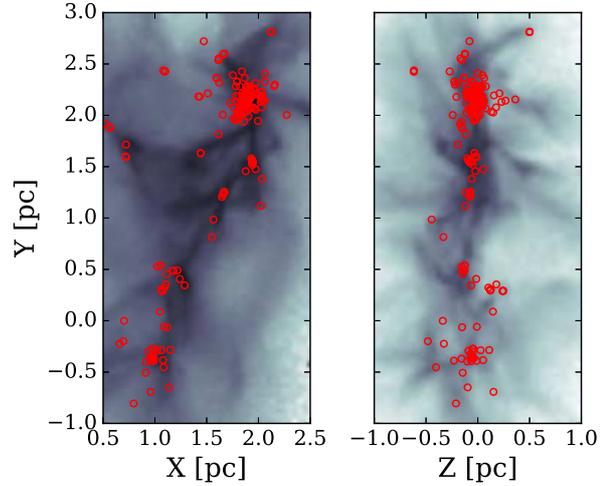}
    \caption{Projection of gas density and sink locations (red circles) at $t = 1.1 \mathrm{t_{ff}}$ for a) XY projection, where the line of sight is along the initial rotation axis b) YZ projection, with a line of sight perpendicular to the initial rotation axis. Spatially, stars generally follow the gas.}
    \label{fig:morph}
\end{figure}
 
\section{Results}
\par At $1.1 t_{\mathrm{ff}}$, run LR formed 265 sink particles (about $950 M_{\odot}$; the median sink mass was about $1 M_{\odot}$ ). The resulting distribution of gas and sink particles, seen in two distinct projections in Figure \ref{fig:morph}, exhibits the filament and hub morphology seen in many star forming regions (for example, the NH$_3$ filaments in OMC 1; \cite{Wiseman_1998}; see also \cite{Teixeira_2006}, \cite{Peretto_2014}). The projections vary due to the preferential direction of the angular momentum that was initially injected into the system. The left (XY) projection  has the line of sight along the initial rotation axis so that all rotation is in the plane of the image, while the right (YZ) projection's line of sight is perpendicular to the rotation axis. 

The global structure of gas and sinks is
qualitatively similar to that of Orion A
\citep[e.g., Figure 1 of][]{DaRio_2017}. This
is a reflection of the in intial geometry of the cloud, with the ellipsoidal shape resulting in forming a filament and the
concentration of gas and stars resulting
from gravitational focusing near the closed
end of the ellipse \citep[e.g.,][]{Hartmann_2007}.  However, it is the action of gravity that creates these global structures out of an initially smooth distribution, with the sinks and groups forming from gravitational collapse of the substructure introduced by the random initial velocity perturbations.

\begin{figure}
    \centering
    \includegraphics[width = \columnwidth]{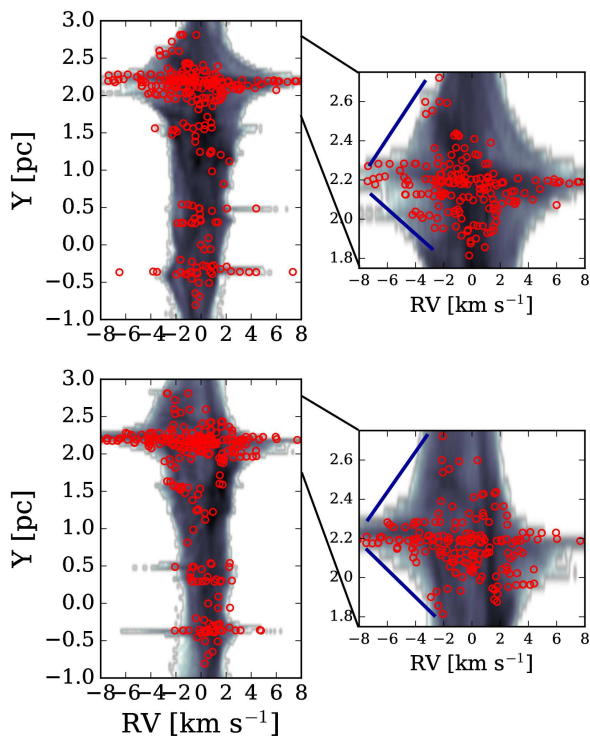}
    \caption{Position-velocity diagrams for the gas and sinks (red circles) at $t = 1.1 \mathrm{t_{ff}}$ for  XY projection (top row), where the line of sight is along the initial rotation axis. The inset is a zoomed in view, annotating the velocity gradients along the gas in the hub cluster region which are on the order of $6 \mathrm{km}$ $\mathrm{s^{-1} pc^{-1}}$.The YZ projection (bottom row) has a line of sight perpendicular to the initial rotation axis. The inset shows these velocity gradients are $4-7 \mathrm{km}$ $\mathrm{s^{-1} pc^{-1}}$. }
    \label{fig:pvs}
\end{figure}

\par The position-radial velocity diagrams for the two projections shown in Figure \ref{fig:pvs}
again show a qualitative resemblance to the
observations of Orion A \citep[Figure 1 of]{DaRio_2017}, with larger velocity dispersions where groups of sinks (stars)
are present, particularly in the major group at $Y \sim 2.3$~pc. These velocity dispersions are simply the result of gravitational acceleration as the density concentrations collapse. In addition, the gas shows 
V-shaped velocity gradients centered on the ONC-like main cluster, as observed in OMC-1 by
\citet{Hacar_2017} (see \S 4).
The main difference between the qualitative
behavior of our simulation and the observations of Orion A
is that our simulation does not show as large overall velocity gradient
from top to bottom.

\begin{figure*}
    \centering
    \includegraphics[width = 2\columnwidth]{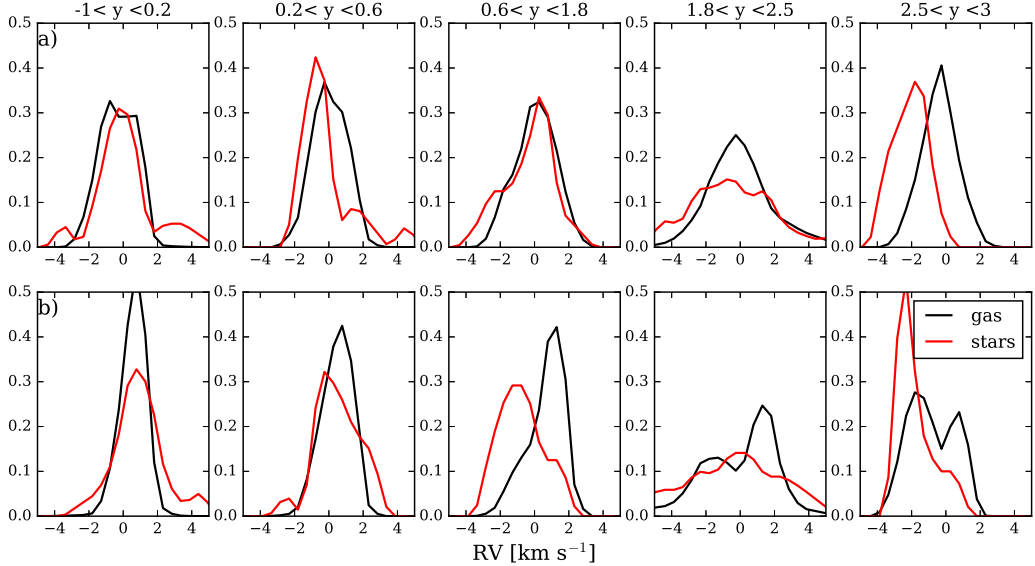}
    \caption{ Probability density functions of radial velocities for the stars and gas,
    taken in slices perpendicular to the long axis, shown for two different projections: a) XY (top row) and b) YZ (bottom row). Each column corresponds to a slice at a position along the long axis of the filament containing a grouping of stars. The first 3 columns are populations of stars along the south end of filament, while the 4th column is the slice of the main ONC-like hub cluster region. The last column is a smaller sub cluster region north of the main cluster.}
    \label{fig:rvs}
\end{figure*}

The optical radial velocity survey of \citet{Tobin_2009} found that while the stars mostly follow the motions of the dense
($^{13}$CO) gas, there is a broader wing of members with more blueshifted velocities.
This issue was reinvestigated by \citet{DaRio_2017} using their near-infrared
radial velocity measurements, and thus with a sample less affected by extinction.  They similarly found a blue offset due to an asymmetry in the stellar velocity distribution with respect to the CO emission.  To explore
our simulation in this respect, we constructed probability density distributions of velocities in slices perpendicular
to the main filament at different $Y$ positions, as in Figure 4 of \citet{DaRio_2017}.  Figure \ref{fig:rvs} shows that, as expected, most of the sinks (stars) follow the gas, although in some 
projections and certain positions we find
a small (1 - 2 \, ${\rm km \, s^{-1}}$)
blueshift.  It is not obvious that this
shifted population is simply that of
a broad wing, although this conclusion is tentative given our small number
statistics.  The origin of the blueshift is
not clear, but it may be a result of the
initial rotation of the cloud, given that it is more evident in the $YZ$ projection.

The velocity dispersions of stellar groups are also larger than that of the gas, as also seen in Orion A \citep[e.g., Figure 10 in][]{DaRio_2017}.
This is especially evident in groups at smaller $Y$. The larger stellar dispersions are the result of close interactions between sinks.

\par
\begin{figure*}
    \centering
    \includegraphics[width = 2\columnwidth]{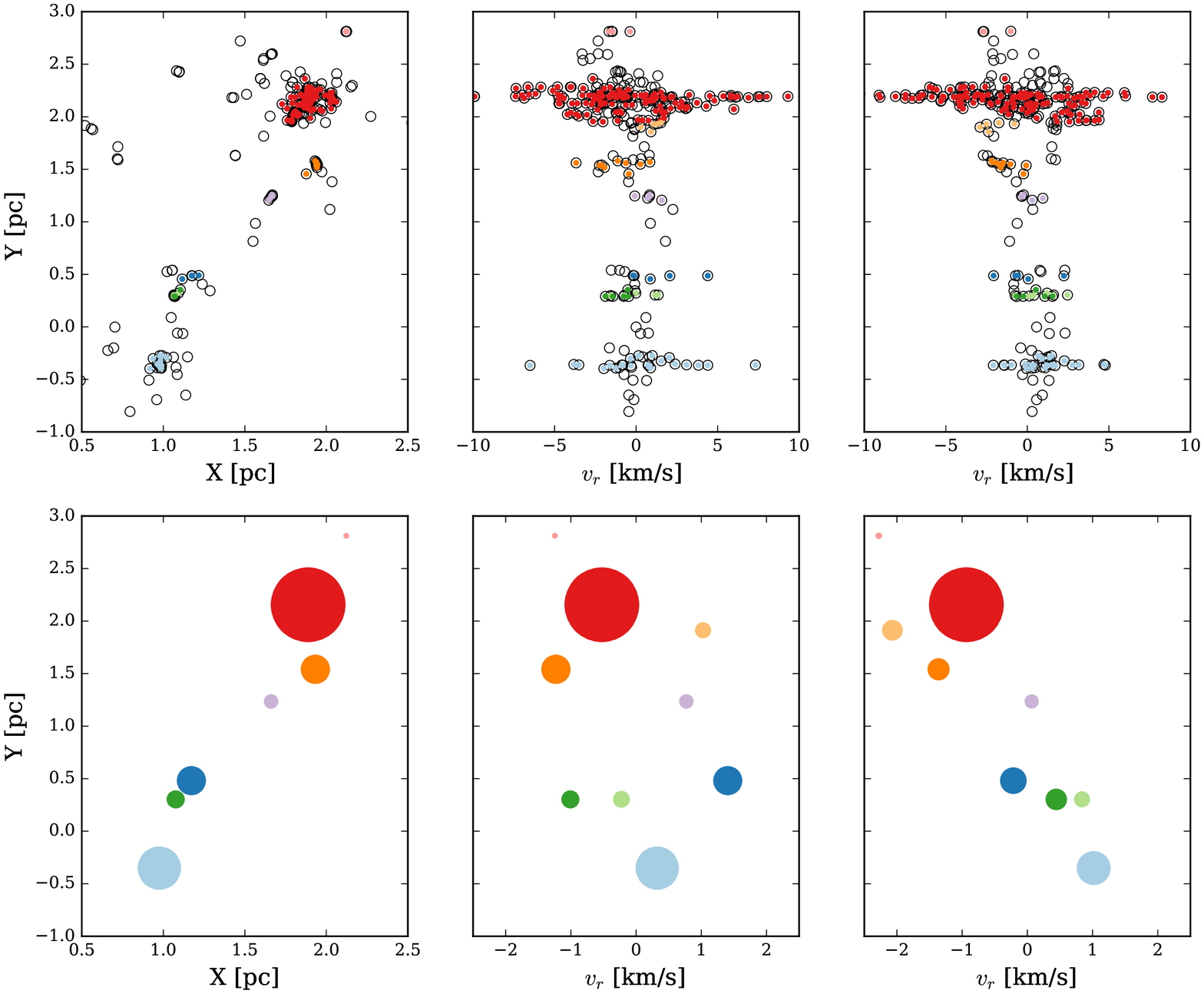}
    \caption{Kinematic substructure identified by FOF at $t=1.1 \mathrm{t_{ff}}$. The ONC-like cluster is the largest group with N = 128 members out of N = 265 total stars.} Top row: position and position-velocity diagrams for individual stars. Colored circles represent stars that have been assigned to groups. Bottom row: position and position-velocity diagrams for stellar substructures. Circles are centered around the mean velocity of the identified structures and their sizes correspond to their relative projected areas in position-position-velocity space. The middle column is the position-velocity diagram in the XY projection and the rightmost column is the YZ projection. 
    \label{fig:substructure}
\end{figure*}

 We identify kinematic substructure by the method described in \S 2 at different times during the simulation. At early times, the substructures are representative of the early subclusters, collapsed cores along the two main filaments which form by early collapse along the boundary of the ellipsoid. During the course of collapse, kinematic substructures grow larger via two mechanisms: by the addition of stars forming within their potential wells and by merging with other substructures. By the end of the simulation at $t=1.1 \mathrm{t_{ff}}$, substructures in the stars trace out the main cluster and the densest parts of the largest filament (Figure \ref{fig:substructure}).  The resulting groups, which are the product of clustering due to gravitational interactions in the subvirial cloud, qualitatively represent the
PPV groups in Orion A identified by
\citet{DaRio_2017} (their Figure 7).

\section{Discussion}
\par In the context of cold collapse, we find that filaments represent bound structures that can form contemporaneously with other structures of high stellar density. While \citet{Burkert_2004} have shown that gravitational collapse tends occur along the shortest dimension first, proceeding from cloud to sheet to filaments to cores, gravitational collapse also amplifies any anisotropies, meaning this process happens at a multitude of scales and speeds such that all of these structures form at similar times. Gravitational focussing can create a gravitational potential minimum at locations of high curvature, such that, unlike in \citet{Myers_2009}, the creation of both a hub and filaments does not require a spherical overdensity in the location of the hub. This also means that the familiar hub-filament paradigm does not necessitate that filaments form first or even that filaments form the hub by feeding it material, although flow along filaments can occur as a result of the hub usually representing a potential minimum, like rivers flowing downstream into the ocean. 
\par While the parameters of our simulation were chosen to obtain an overall morphology akin to that of Orion A, there was no intention of producing an exact correspondence. The simulations only incorporate two forces: gravity and pressure gradients so that the simulation can be rescaled with length, mass, and time units such that $L^3 M^{-1} T^{-2} = $ constant. While the precise value of the sound speed is not important since the cloud is highly subvirial thermally,
a rescaling of
the sound speed by $(M/L)^{1/2}$
or equivalently $L/T$
keeps the ratio of the sound crossing time to the free-fall time the same, maintaining
the  relative
importance of pressure to gravity.

\par Because the proportion of mass in our main cluster is larger relative to the entire cloud than the ONC is to Orion A, a single rescaling is not possible to address the entire region. However, the gravitational potential in the main cluster region at the end of the simulation is much deeper than the global cloud (Figure 3 of \citet{Kuznetsova_2015}), so to first order one can treat the cluster dynamics separately from the large scale cloud structure. 

\par At the end of the simulation, the main cluster has a mass of about 500 $M_{\odot}$ (dominated by stars/sinks) within a radius of about 0.2 pc (see Figure 14 of \citet{Kuznetsova_2015}). From Figure \ref{fig:rvs}, the average 1-D velocity dispersion of stars within 0.2 pc of the cluster center is 2.7 km s$^{-1}$, ranging between 2.2 and 3.2 km s$^{-1}$. These velocity dispersions are roughly "virial", consistent with the 1-D estimate assuming isotropic motions $v \sim {3}^{-1/2}\sqrt{2GM/R} \sim  2.7 $ km $s^{-1}$. 

To compare our results to the observed velocity dispersion of the ONC
$\sim$ 2- 2.5 km s$^{-1}$
\citep{Jones_1988,Tobin_2009,DaRio_2017} we need to scale the mass and radius.
The ONC has a mass of $\sim 2000 M_{\odot}$ within a radius of 2 pc \citep{Hillenbrand_1998} in stars alone.The mass in gas within the same region is  uncertain, but \citep{Hillenbrand_1998} argue
that it might be as large as twice the mass in stars. If we adopt a
total ONC region mass of 6000 $M_{\odot}$, we would need to
scale up the cluster region mass
by a factors 12 and the radius by a factor of 10 from 0.2 to 2 pc,
increasing our
velocity dispersion from
$\sim 2.7$ to $\sim$ 3 km s$^{-1}$.
This is easily within the 
uncertainties; if we instead use
the
virial mass estimated by
\citep{Hillenbrand_1998} of
4500 $M_{\odot}$, the resulting
scaled velocity dispersion is
2.5 km s$^{-1}$.

\par We qualitatively observe the evidence of gravitational acceleration toward the cluster center as V-shaped radial velocity gradients centered around the hub cluster (Figure \ref{fig:pvs}),
similarly to the $5-7 \mathrm{km}$ $\mathrm{s^{-1} pc^{-1}}$ V-structure seen
in N$_2$H$^+$ emission from OMC-1 
by \citet{Hacar_2017}. The simulation's symmetric collapse produces two V-shaped features, in the red and the blue, compared to the OMC's one sided feature. A rescaling according to size and mass scales of the ONC can produce velocity gradients consistent with those in OMC-1, but direct comparisons are difficult to do as it depends on the inclination of the filament to the line of sight, which is highly uncertain.

\par The kinematics of the subclusters that form along the filament at the same time as the early natal cluster stars present an interesting opportunity to study the dynamics of star formation without the dynamical processing that occurs in the populated main cluster erasing kinematic signatures. Recent studies of the \emph{APOGEE} data for Orion have identified kinematic substructures \citep{DaRio_2017, Hacar_2016}. While \citet{DaRio_2017} interpret the substructures they find along the filament to have inherited the properties of their natal turbulence, \citet{Hacar_2016} similarly posit that these stellar groupings inherit the dynamical properties of their natal gas clumps. However, we show in the context of our simulations that since these substructures clearly accrete both stars and other subclusters of stars, any signature of natal turbulence will be short lived.  The properties of these structures are more likely determined by the material they accrete; the initial turbulent fluctuations serve only as seeds for the structures to develop. As discussed in detail  in \citet{Kuznetsova_2015}, it is possible for initial infall signatures like proper motions to be preserved by the substructures in the filament. With Gaia and complementary radial velocity studies in the optical and IR \citep[e.g.][]{Tobin_2009, Kounkel_2016}, these filament populations can become an important testing ground for star formation theories.

\par As stellar substructures grow in membership and mass, their velocity dispersions also grow, outpacing the velocity dispersion of the gas and mirroring the global trend identified in \citet{Kuznetsova_2015} for stellar velocity dispersions to be higher than that of the gas. Massive substructures from earlier timesteps in the simulation have smaller dispersions than massive substructures at later times. This further suggests that gravitational interactions between stars during accretion of members could play a role in determining the dynamics of stellar substructures.

\citet{DaRio_2017} showed that in the southern region of Orion A (L 1641), the velocity
dispersion predicted from the estimated gravitational potential of the broad (several pc-wide) filament predicted a velocity dispersion of only about $0.65 {\rm \, km\, s^{-1}}$, much smaller than the observed
dispersions of gas and stars which are on the order of $2.5 {\rm \, km\, s^{-1}}$. They concluded that this region could be gravitationally
unbound. However, the mass to length ratio is a metric reserved for narrow cylindrical filaments, which is not true of either simulated filament or Orion A, which are both 3D objects. While the  velocity dispersions of the simulation substructures also exceed the 1-D filament prediction, the groups are, in fact, bound. The filamentary condition for virial equilibrium does not account for the total mass in gas present deepening the global gravitational potential or for the mass in stars present in the groups, which dominate the local gravitational potential of the sub-clusters.

While the sinks (stars)
constitute only a small fraction of the
gas mass in the region, they are much
more centrally concentrated than the gas,
and so increase the magnitude of the {\em local} gravitational potential over that of the gas \citep[see][]{Kuznetsova_2015}.

\par In embedded systems such as Orion, taking stock of gas mass for properly diagnosing kinematics becomes especially important. In \citet{Kuznetsova_2015}, we show that without complete information about the gas content, the virial parameter, $\alpha = 5\sigma^2 R/GM $, for a gas rich cluster is at best an upper limit for which completely bound systems can appear to be supervirial. The virial parameter assumes that the cluster has the gravitational potential of a uniform sphere and exhibits isotropic motion, neither of which are obviously applicable in Orion, nor in our simulation. We show in \citet{Kuznetsova_2015} that calculating a virial parameter for the cluster in the simulation can yield a range of values from 0.9 to 4.0, depending on how much gas mass is accounted for and how the bounds of the cluster are defined. Thus, the virial parameter of 1.8 from \citet{DaRio_2017} could easily result from underestimating the mass of the system.

\par We have shown that stellar groups can appear red or blue shifted from the bulk motions of the gas, depending on the projection used (Figure \ref{fig:rvs} ). It is also evident from Figure \ref{fig:rvs} that projection effects can hide or display the presence of rotation. The projection dominated by the rotational component (YZ) shows the clear development of a double peaked profile from north to south along the filament. This projection also most closely matches the observed velocity gradient of substructures in Orion A shown in \citet{DaRio_2017}, but attributing that phenomenon to angular momentum injection at large scales is difficult. The gas in the Northern end of Orion A has likely been disturbed by the formation of super bubbles \citep{Bally_2008}, corrupting velocity information that could definitively identify large scale rotation. However, velocity gradients themselves are susceptible to projection effects where the projection of filament populations along the line of sight to the cluster can produce a radial velocity gradient in stars and gas along the filament axis \citep{Kuznetsova_2015}. Even without projection, the subcluster populations along the filament can be expected to  have different velocities from the hub cluster simply because they are kinematically different, having had a separate dynamical history than stars that end up processed by the hub cluster.
 
\par 
We emphasize that the only way
in which we "seeded" the initial condition in a way to produce a desired outcome was in the shape of the initial ellipsoidal cloud, which provided an axis of symmetry to produce somewhat filamentary structure, along with a smaller radius of curvature at one end which the simulations of \citet{Burkert_2004} and \citet{Hartmann_2007} showed would produce a focal point to form a massive region and hence a large cluster.  The other groups
of stars arose from the
density perturbations resulting from the
random initial velocity fluctuations.

The results of this simple simulation
illustrate the power of gravity to naturally produce groups and clusters of stars. This is in line with our previous
study in which we argued that gravitational focusing effects are
responsible for producing the observed power-law mass distribution of young clusters \citep{Kuznetsova_2017}.

\section{Summary}
\par We perform a new analysis of simulation datasets from \citet{Kuznetsova_2015} to identify and characterize the type of substructures that can be expected in a star forming region created by the global gravitational collapse of molecular gas. We find that substructure such as filaments and sub cluster forms contemporaneously with many of the stars that end up in the main star cluster. These structures could be useful probes of star cluster formation at early times and complete kinematic information could provide a detailed test of star formation theories. 
\par In addition, we examine and compare the kinematic features identified in Orion A by \citet{DaRio_2017} to those we can identify in our simulations. We find that features, such as the difference between stellar and gas velocity dispersions in groups of stars, are reproduced by our simulations. Disparities between the kinematics of filament and cluster stars could be due to rotation or manifest primarily due to projection and/or extinction effects, which should always be considered when dealing with anisotropic dense structures.  Finally, we find a V-shaped
structure in the gas around our most massive
cluster very similar to that observed by
\citet{Hacar_2017} in the ONC region,
supporting their interpretation of
collapse.

Our simple cold-collapse model driven entirely by gravity can easily produce the familiar hub-filament morphology such that turbulence is not required to play a leading role in forming substructures. While simulations including many other 
effects, such as magnetic fields and stellar
feedback, will ultimately be required to
fully understand the structure of
star-forming regions such as Orion, our
calculations focusing on the effects of
gravity can provide an initial basis for
more complicated simulations in the future.

\section{Acknowledgments}

AK acknowledges funding from the Rackham Graduate School of the University of Michigan and from NASA grant NNX16AB46G. J.B.-P. acknowledges UNAM-PAPIIT grant number IN110816, and to UNAM's DGAPA-PASPA Sabbatical program. He also is indebted to the Alexander von Humboldt Stiftung for its invaluable support. Numerical simulations were performed at the Miztli cluster at DGTIC-UNAM.

\bibliographystyle{mnras}
\bibliography{biblio}

\bsp	
\label{lastpage}
\end{document}